\documentclass[twocolumn]{aastex63}
\submitjournal{ApJL}

\shorttitle{GW Dissipation in Gravitationally Bound Systems}
\shortauthors{Loeb}

\begin{document}

\title{Upper Limit on the Dissipation of Gravitational Waves in
  Gravitationally Bound Systems}

\correspondingauthor{Abraham Loeb}
\email{aloeb@cfa.harvard.edu}

\author{Abraham Loeb}
\affiliation{Astronomy Department, Harvard University, 60 Garden Street,
  Cambridge, MA 02138, USA}

\begin{abstract} 

It is shown that a gravitationally bound system with a one-dimensional
velocity dispersion $\sigma$ can at most dissipate a fraction $\sim 36
(\sigma/c)^3$ of the gravitational wave energy propagating through it,
even if their dynamical time is shorter than the wave period. The
limit is saturated for low frequency waves propagating through a
system of particles with a mean-free-path equal to the size of the
system, such as hot protons in galaxy clusters, strongly-interacting
dark matter particles in halos, or massive black holes in
clusters. For such systems with random motions and no resonances, the
dissipated fraction, $\lesssim 10^{-6}$, does not degrade the use of
gravitational waves as cosmological probes. At high wave
frequencies, the dissipated fraction is additionally suppressed by the
square of the ratio between the collision frequency and the wave
frequency.  The electromagnetic counterparts that result from the
dissipation are too faint to be detectable at cosmological distances.

\end{abstract}


\section{Introduction} \label{sec:intro}

The discovery of gravitational waves (GWs) by LIGO \citep{LIGO2016}
revolutionized observational astronomy by expanding its means for
detecting sources at cosmological distances beyond electromagnetic
radiation \citep{Metzger2019}. In particular, GW sources with known
redshifts can serve as ``standard sirens''
\citep{Schutz1986,Holz2005,Chen2019} for measuring accurately 
cosmological distances, while avoiding the uncertainties or systematics
of traditional ``distance ladder'' techniques
\citep{Riess2019,Freedman2019,Verde2019,Foley2020}, since the GW
source physics is well understood.

An implicit assumption in all past discussions on measuring
cosmological distances with GW sources is that the GW signal is not
modified as it propagates through intervening matter. This constitutes
a key advantage of GWs relative to ``standardized candles'' of
electromagnetic radiation, such as Type Ia supernovae
\citep{Riess2019}, which could be absorbed by intervening gas and dust
along the line-of-sight \citep{Aguirre1999}.

Nevertheless, a medium with a dynamic viscosity coefficient $\eta$ could
dissipate the energy density of GWs on a dissipation timescale
\citep{Hawking1966,Weinberg1972},
\begin{equation}
t_{\rm diss}={c^2\over 16\pi G\eta},
\label{tdiss}
\end{equation}
where $c$ is the speed of light and $G$ is Newton's constant. To
within a factor of order unity, the dynamic viscosity coefficient can
be expressed as \citep{Chapman1970},
\begin{equation}
\eta\sim \rho \lambda \sigma ,
\label{visc}
\end{equation}
where $\lambda=1/(nA)$ is the collision mean-free-path, $n=(\rho/m)$
is the particle number density corresponding to a mass density $\rho$,
$A$ is the collision cross-section, $\sigma\equiv {1\over 3} \langle
v^2\rangle^{1/2}$ is the one-dimensional velocity dispersion, and $m$
is the mass of the particles that make up the dissipative
medium. Equations (\ref{tdiss}-\ref{visc}) hold as long as the GW
period is larger than the system's dynamical time, so that the
particles behave as a fluid during the passage of the GW. 

Other effects, such as resonances
\citep{McKernan2014,Annulli2018,Montani2019,Servin2001}, could enhance
the dissipation even in collisionless systems.  In particular, the
cosmic neutrino background dissipated the energy of primordial GWs by
up to $35.6\%$ for comoving wavelengths that entered the horizon
during the radiation dominated epoch \citep{Weinberg2004}.  One may
wonder whether GW dissipation would also be significant also in the
dense environments of galactic nuclei, where some GW sources are
preferentially formed \citep{Loeb2010,Bartos2017,Tagawa2019}, even if
environmental heating of stars or accretion disks by GW sources is not
sufficiently strong to be detectable at extragalactic distances
\citep{KL2008,Li2012}. For simplicity, we focus on systems with random
motions and no resonances.

As long as the GW period is longer than the system's dynamical time,
the dissipation time is minimized for a system with a radius, $R$,
that is comparable to the collision mean-free-path of its particles,
$\lambda$. Shorter mean-free-paths result in a smaller viscosity
coefficient and longer values are not allowed since the particles are
confined to the system.  Collision rates below the optimal value only
reduce the level of dissipation during the passage of the GW through
the system.

Examples for optimal systems with $\lambda\sim R$ include hot protons
in clusters of galaxies \citep{Loeb2007}, strongly-interacting dark
matter in halos \citep{Goswami2017,Fitts2019}, and massive black holes
that scatter off each other gravitationally in
clusters\footnote{Interestingly, black holes with masses of order
  $M_{\rm BH}\sim 10^{5}M_\odot$ possess a cross-section per unit mass
  for gravitational scattering off each other, $(A/m)\sim\pi (GM_{\rm
    BH}/\sigma^2)^2/M_{\rm BH}$, which overlaps with the value of
  $(A/m)\sim 1~{\rm cm^2/g}$ needed to alleviate the cusp-core problem
  in dwarf galaxies, as it provides $\lambda\sim R$ at relative speeds
  of $\sigma \sim 10~{\rm km~s^{-1}}$. The velocity scaling,
  $(A/m)\propto \sigma^{-4}$, reduces the collisional effect in more
  massive halos, as envisioned for dark matter with a Yukawa potential
  \citep{Loeb2011}. Unfortunately, massive black holes cannot serve as
  primary candidates for strongly interacting dark matter based on
  other constraints \citep{Carr2019}. But a cluster of them can
  dissipate GW energy by converting it into an increase in $\sigma$
  (``heat'') through two-body scatterings.}.  The maximal dissipation
in these examples would be achieved for primordial GWs of very low
frequencies, $\lesssim (\sigma/R)=1/t_{\rm dyn}$, or for GWs produced
by binaries with an orbital period longer than the dynamical time of
the absorbing system, $t_{\rm dyn}$.

This brief note sets an upper limit on the level of dissipation that a
GW signal encounters by passing through astrophysical systems that are
bound by gravity for arbitrary GW frequencies. The limit is
independent of the composition or nature of the absorbing medium as
long as there are no resonances with the GW frequency
\citep{McKernan2014,Annulli2018,Montani2019,Servin2001}.  Its general
validity clears the way for using GW sources for precise cosmological
measurements by observatories such as
LIGO/Virgo\footnote{https://www.ligo.org/},
LISA\footnote{https://www.elisascience.org/} or their future
extensions \citep{Hall2019}.

The frequency-independent expressions (\ref{tdiss}-\ref{visc}) are
valid as long as the GW frequency is smaller than the collision
frequency of particles in the system. Otherwise, dissipation is
suppressed since particles have a low collision probability per GW
period, after which they return to their original position and
velocity with no memory of previous oscillations. In this high GW
frequency regime, the velocity shear being dissipated is dictated by
the amplitude of the periodic motion of the particles. We derive this
additional (frequency-dependent) suppression of the GW dissipation in
the concluding section.

\section{Absolute Dissipation Limit for Arbitrary GW Frequency}

Let us consider a collisional system of radius $R$, composed of
particles that are bound by gravity, without making any assumptions
about the nature of the constituent particles. A GW signal would cross
the system over a timescale $t_{\rm cross}\sim (R/c)$, which is
shorter than the crossing-time by the system particles, $\sim
(R/\sigma)$. During the GW passage, viscous dissipation is maximized
for $\lambda\sim R$, as already noted. Larger values of the
mean-free-path, $\lambda$, are not allowed because particle
trajectories are gravitationally confined to the system size.  Smaller
values of $\lambda$ reduce the viscosity coefficient based on equation
(\ref{visc}).

The fraction of the GW energy which the system absorbs is,
\begin{equation}
\epsilon_{\rm diss}\sim {t_{\rm cross}\over t_{\rm diss}}.
\label{ratio}
\end{equation}
Substituting the maximum viscosity coefficient, $\eta\sim \rho R
\sigma$, into equation (\ref{tdiss}), yields an upper limit on the
dissipated fraction of the GW energy,
\begin{equation}
\epsilon_{\rm diss} < \left({16\pi}\right) {G\rho R^2\sigma\over c^3}.
\label{epslim}
\end{equation}

For a system bound by the gravitational potential of the dissipating
particles plus other components, such as gas, stars, black holes or
dark matter, the {\it Virial Theorem} implies \citep{Binney2008},
\begin{equation}
\left({4\pi\over 3}\right) G\rho R^2 < 3 \sigma^2,
\label{mlim}
\end{equation}
where the inequality stems from the fact that the dissipating particles
with a mean mass density $\rho=M(<R)/[(4\pi/3)R^3]$, account for only
a fraction of the total mass density in the system which could include
additional components. Substituting (\ref{mlim}) into (\ref{epslim})
yields our final upper limit:
\begin{equation}
\epsilon_{\rm diss}< 36\left({\sigma\over c}\right)^3 .
\label{final}
\end{equation}
The numerical coefficient on the right-hand-side of (\ref{final})
could change by a factor of order unity, depending on the detailed
radial profile of $\rho$ and $\lambda$ within the system.

\section{Electromagnetic Counterparts}

The final result (\ref{final}) implies that the fraction of GW energy
that can be absorbed by any self-gravitating system of a
one-dimensional velocity dispersion $\sigma$ is limited to $\sim 36
(\sigma/c)^3$. Dark matter halos possess a maximum value of
$(\sigma/c)\lesssim 10^{-2.5}$ in clusters of galaxies
\citep{Loeb2013}, and cannot dissipate more than $\sim 10^{-6}$ of the
the GW energy from a source hosted by them or located behind
them. This limit applies to all possible values of the
self-interaction cross-section per unit mass of dark matter particles
at all GW frequencies.

The negative heat capacity of gravitationally-bound systems makes them
vulnerable to the gravothermal instability
\citep{Balberg2002,Hennawi2002}. As a result, compact systems with
large values of $(\sigma/c)$ and a short collisional mean-free-path,
$\lambda \ll R$, could evolve to a black hole or evaporate on a
timescale shorter than the age of the Universe.

Consequently, the amplitude of GW signals cannot be absorbed by
intervening gravitationally-bound systems to any significant level
that would degrade their potential use for cosmology
\citep{Schutz1986,Holz2005,Chen2019}. In particular, uncertainties in
the peculiar velocities of GW sources are of order $\sim \sigma/c$ and
exceed by a factor $\gtrsim 0.03 (c/\sigma)^2$ the level of viscous
dissipation within their host dynamical system.

The above results also limit a possible electromagnetic (EM)
counterpart to the GW signal from its environment
\citep{KL2008,Li2012}, unrelated to the possible EM emission by the
source itself \citep{Loeb2016,DOrazio2018,Metzger2019}. The
dissipation of a fraction $\epsilon_{\rm diss}$ of the GW energy,
$E_{\rm GW}$, in a baryonic system surrounding the GW source, would
lead to an EM counterpart with a luminosity,
\begin{equation}
L_{\rm EM}\sim \epsilon_{\rm cool}\epsilon_{\rm diss} \left({E_{\rm GW}\over
t_{\rm cool}}\right),
\label{LEM}
\end{equation}
where $\epsilon_{\rm cool}$ is the fraction of the dissipated energy
that gets radiated electromagnetically over a cooling time, $t_{\rm
  cool}$. The time delay across the system sets a lower limit on the
cooling time, $t_{\rm cool}\gtrsim (R/c)$, and hence an upper limit on
the EM luminosity based on (\ref{final}) and (\ref{LEM}) for the
ultimate radiative efficiency of $\epsilon_{\rm cool}\sim 1$,
\begin{equation}
L_{\rm EM,max}\lesssim 10^{40}~{\rm {erg\over s}}\left({E_{\rm GW}\over
  0.1M_\odot c^2}\right)\left({\sigma\over
  10^{-3}c}\right)^3\left({R\over 0.01~{\rm pc}}\right)^{-1},
\label{EMmax}
\end{equation}
over a period of $\gtrsim 10~{\rm days}~(R/0.01~{\rm pc})$.
The limit is nearly twenty orders of magnitude below the maximum
attainable GW luminosity, $\sim (c^5/G)=4\times 10^{59}~{\rm
  erg~s^{-1}}$. It can also be normalized by the Eddington EM limit
for the total mass $M_{\rm tot}$ of the host dynamical system, $L_{\rm
  Edd}=1.4\times 10^{44}~{\rm erg~s^{-1}}(M_{\rm tot}/10^6M_\odot)$
\citep{Loeb2013}.  Using the {\it Virial Theorem} again, $(GM_{\rm
  tot}/R)\sim 3\sigma^2$, the normalized upper limit is tight,
\begin{equation}
{L_{\rm EM,max}\over L_{\rm Edd}}\lesssim 10^{-4} \left({E_{\rm
    GW}\over 0.1M_\odot c^2}\right)\left({\sigma\over
  10^{-3}c}\right)\left({R\over 0.01~{\rm pc}}\right)^{-2},
\end{equation}
implying that EM counterparts from viscous dissipation of GW signals
at cosmological distances are too faint to be detectable by existing
telescopes.

\section{Further Suppression at High GW Frequencies}

The above limits were derived without a reference to the GW frequency.
However, the frequency-independent dissipation rate in equation
(\ref{tdiss}) could be amplified by resonances of the GW frequency
with modes in the medium, such as those associated with binary systems
\citep{McKernan2014,Annulli2018,Montani2019} or a magnetic field
\citep{Servin2001}.

In thermal systems with random motions of particles, the
standard viscous dissipation rate increases with increasing
mean-free-time between collisions because particles are able to sample
a steadily increasing velocity offset in the underlying shear
flow. Since the GW-induced shear reverses sign on the GW period, the
fact that a particle waits longer than a wave period for the next
collision does not help it develop more velocity offset relative to
the local flow. The maximum shear it samples is the value that the GW
induces over a single wave period, and this fixed amount is dissipated
over the collision period. This is in contrast to the behavior at
short collision periods where the shear sampled is inversely
proportional to the collision period, yielding a dissipation rate in
(\ref{tdiss}) that is proportional to the viscosity
coefficient\footnote{Note that the collision period must be compared
  relative to the GW period rather than the mean-free-path relative to
  the GW wavelength. The two criteria are different for
  non-relativistic particles.}.
  
Equation (\ref{final}) provides the absolute upper limit on
$\epsilon_{\rm diss}$ for any GW frequency by considering the maximum
possible value of $\eta$, but the actual limit on $\epsilon_{\rm
  diss}$ at high GW frequencies is tighter by the square of the ratio
between the lower collision frequency and the GW frequency. This can
be derived as follows.

In analogy with the propagation of EM waves in a collisional plasma
\citep{Braginski1965,Stix1992}, the introduction of a {\it Crook
  collision term}, $-\nu_{\rm coll}{\bf v}$, to the momentum equation
describing the acceleration of a particle, $d{\bf v}/dt$, by a GW that
oscillates over time $t$ as $\propto e^{i\omega t}$ with a GW
frequency $\omega$, leads to a dissipation rate that rises inversely
with the frequency ratio,
\begin{equation}
f_{\rm ratio}=\left({\nu_{\rm coll}\over\omega}\right) ,
 \end{equation} 
 for $f_{\rm ratio}\ll 1$ [consistently with equation (\ref{tdiss})],
 peaks at $f_{\rm ratio} \sim 1$ and then declines in proportion to
 $f_{\rm ratio}^{-1}$ for $f_{\rm ratio}\gg 1$. 

At high GW frequencies, the above formulation provides an additional
(frequency-dependent) suppression factor of,
\begin{equation}
{1\over 1+f_{\rm ratio}^{-2}},
\end{equation}
on the right-hand-side of the limit (\ref{final}). This suppression
factor does not depend on the nature of the periodic driving force (be
it EM or GW) but only on the collisional dynamics of the particles in
the system. This extra suppression factor obtains extremely small
values at the GW frequencies detectable by LIGO and most dissipating
systems.

In the transition regime, where $f_{\rm ratio}\sim 1$, the dissipated
fraction is limited by,
\begin{equation}
\epsilon_{\rm diss}<36\left({\sigma\over c}\right)^3 \left({1\over
  \omega t_{\rm dyn}}\right).
\end{equation}
where we have used equations (\ref{tdiss})-(\ref{ratio})
and (\ref{mlim}), and the relations $t_{\rm dyn}=(R/\sigma)$ and
$\nu_{\rm coll}=(\sigma/\lambda)$. This limit applies only for $\omega
t_{\rm dyn}\gtrsim 1$, with the limit (\ref{final}) being saturated at
lower frequencies.

\acknowledgements

My special thanks go to Bence Kocsis and Martin Rees for inspiring
correspondences that led to the addition of the last section on high
GW frequencies.  After the completion of this addition, I received a
research note from Jeremy Goodman through the AAS Editorial office, in
which the excess suppression at high GW frequencies was derived
independently based on an elegant mathematical formalism. I am deeply
grateful to Jeremy for his interest in my manuscript and for
communicating his insightful derivation.  This work was supported in
part by the {\it Black Hole Initiative} at Harvard University, which
is funded by grants from GBMF and JTF.

\bibliography{p}{}
\bibliographystyle{aasjournal}

\end{document}